\documentclass[aps,prb,amsmath,amssymb]{revtex4-1}
\usepackage{graphicx}
\usepackage{epsfig,rotate,latexsym}
\usepackage{color, bm}
\usepackage{dashrule}
\begin{document}
\title {Anisotropic super-paramagnetism in cobalt implanted rutile-TiO$_{2}$ single crystals}
\author{Shalik Ram Joshi$^1$, B. Padmanabhan$^2$, Anupama Chanda$^3$, N. Shukla$^4$, Vivek Malik$^2$,
D. Kanjilal$^5$ and Shikha Varma$^1$}
\email{shikha@iopb.res.in}
\affiliation{$^1$Institute of Physics, Sachivalaya Marg, Bhubaneswar-751005\\ 
$^2$Department of Physics, Indian Institute of Technology, Roorkie 247667, India\\
$^3$Department of Physics, Dr. Hari Singh Gour University, M.P 470003, India\\
$^4$Department of Physics, NIT Patna, 800005, India\\
$^5$Inter University Accelerator Center, New Delhi 110067, India}
\date{\today}
\begin{abstract}
We study the magnetic properties of single crystals of rutile-TiO$_{2}$ implanted with cobalt for various fluences ranging from 1x10$^{16}$ to 1x10$^{17}$. For lower
fluences, the nature of magnetism is not affected by the Co substituting Ti sites forming Ti$_{1-x}$Co$_{x}$O$_{2}$. From a fluence of 5x10$^{16}$ onwards, the magnetic
behaviour changes drastically due to formation of hcp cobalt clusters which are detected by x-ray diffraction. For the highest fluence sample (1x10$^{17}$) we also
detect formation of CoTiO$_{3}$ phase. The Co nano-clusters give rise to super-paramagnetism which is highly anisotropic w.r.t the crytallographic directions of
TiO$_{2}$. The temperature and field dependent magnetization was studied in detail for magnetic field ($H$) along $<$001$>$ and $<$1${\bar 1}$0$>$ directions.
The temperature variation of zero field cooled (ZFC) and field cooled (FC) magnetization shows a much higher blocking temperature ($T_{B}$) along $<$1${\bar 1}$0$>$.
Similarly the scaling of magnetization isotherms above $T_{B}$ is seen only when the field is parallel to $<$1${\bar 1}$0$>$ direction. With field along this direction,
the magnetization shows near saturation at a much smaller field compared to that of$<$001$>$  direction. With increase in fluence we find increase in particle size
(from 2.5 to 4.5 nm diameter) and size distribution by fitting our magnetization isotherms to Langevin function assuming a lognormal distribution of particle sizes.
Below $T_{B}$, at the lowest temperatures we observe that the $M-H$ curves show a wide hysterisis loop when the field is along $<$1${\bar 1}$0$>$ direction
suggesting the highly oriented nature of the clusters. The Co nanoclusters possess an ``easy" and ``hard" axis of magnetization coupled by the magnetocrystalline
anisotropy of Ti$_{1-x}$Co$_{x}$O$_{2}$. In addition, at T=2 K we observe a crossover in the magnetization vs field isotherms between the two field directions in the
samples with Co fluences of 8x10$^{16}$ and 1x10$^{17}$ which is not seen above $T_{B}$. The origin of this crossover, is discussed briefly in terms of anisotropic
paramagnetism arising from cobalt present in 2+ ionic state with $S=3/2$. 
\end{abstract}
\maketitle
\section{Introduction}
\label{introduction} 
The study of magnetism in nano-particles has gained enormous interest in last two decades 
from technological as well as fundamental perspectives\cite{Guimaraes}. In nanoscale systems, magnetic nature
can be drastically different, compared to a bulk, since surface effects play crucial role in determining
this behavior such that properties as diverse as ferromagnetism, anti-ferromagnetism, super-paramagnetism (SPM)
or spin-glass(SG) like behavior are observed \cite{Guimaraes, Kaneyoshi, Bean, Binder}. Among these, 
SPM is a property that crucially depends on the size of the nano-particle and shows magnetic moment that
is proportional to the particle- volume \cite{Bean}. SPM nano-sized cluster can exhibit giant
magnetic moment (sometimes as high as few thousand ${\mu}_{B}$) which is randomly oriented in the
absence of external field. The particles are non-interacting, except for a weak dipole interaction.
Thus, an ideal super-paramagnet should exhibit {\it paramagnetic\ behavior},
i.e follow Curie - Weiss law, but with a large effective moment and a non-hysteretic $M-H$ curve up to 
0~K. However, effects of magneto-crystallinity as well as surface and shape anisotropy alter this 
behavior such that a real super-paramagnet shows deviations from the Curie-Weiss law at the non-zero 
temperatures \cite{Fonseca}. The super-paramagnetic clusters possess
a uniaxial anisotropic direction which is random in direction for each SPM particle. Thus every 
individual nano-particle has its corresponding easy axes of magnetization. 

As the system is cooled through the SPM state, there comes a characteristic temperature called the
blocking temperature ($T_{B}$). Above $T_{B}$, the magnetic moment of the individual SPM particle
is oriented randomly like a normal paramagnet, which can rotate freely under the influence of external field. 
Below $T_{B}$, the individual SPM particle has its magnetic moment blocked along its respective easy anisotropy
axis. This temperature is prominently seen as a bifurcation of zero field cooled (ZFC) and field cooled (FC)
magnetization. Associated with this temperature is an energy barrier $U$, which is the energy required for the
individual magnetic moment to flip its direction along the two easy axes directions. The time for flip is
giving by the characteristic equation \cite{Bean}, 
\begin{equation}
{\tau}= {\tau}_{0}exp\left(\frac{U}{k_{B}T}\right)
\end{equation}
Here $\tau_{0}$ is the limiting relaxation time, $k_B$ is the Boltzmann constant, $T$ is the temperature
and $U$ is the potential barrier. The blocking temperature is affected by individual particle volume
and their distribution \cite{Luis}.

SPM behavior has been shown by several transition metals and their alloys when incorporated in non-magnetic host matrices,
such as by Co multilayers on Al$_{2}$O$_{3}$ \cite{Luis}, nano-crystallites of $CoFe_{2}O_{4}$ \cite{Rondinone}, nano-powdered
CoPt$_{3}$ alloys \cite{Wiekhorst} and Ni nano-particles on SiO$_{2}$ \cite{Fonseca}. These compounds show a near perfect SPM
behavior with a low blocking temperature and a universal scaling behavior \cite{Fonseca} in $M-H$ curves. 

Incorporating Co in TiO$_{2}$, however, has been demonstrated to form a dilute magnetic semiconductor (DMS) 
system that shows room temperature ferromagnetism \cite{Mat}. DMS materials, achieved by introducing 
small concentration of transition metal or non-magnetic material in a semiconductor, exhibit
ferromagnetic/ anti-ferromagnetic properties that are useful in spintronic devices. DMS is
demonstrated  by many systems like Co/Mn substituted in TiO$_{2}$, ZnO or Mn doped in narrow band 
gap semiconductors like GaAs, InAs \cite{Ueda, Fukumura, Ohno1, Edmonds} etc. For Co doped TiO$_2$,
magnetic properties have been investigated, in rutile and anatase forms of bulk as well as 
thin TiO$_2$ \cite{Mat, Kim, Kim2,Toyosaki} films that were prepared by various methods 
like pulsed laser deposition, molecular beam epitaxy, magnetron sputtering, metal organic 
chemical-vapor deposition and sol-gel technique \cite{Mat2, Xu}. In spite of extensive 
studies, origin of observed ferromagnetism in this system is still unclear. 
Investigations suggest that cobalt ions in thin TiO$_{2}$ films exist in a +2 oxidation state 
forming Ti$_{1-x}$Co$_{x}$O$_{2}$, which is ferromagnetic in nature \cite{KimD}. Most of the 
preparation techniques produce precipitation of cobalt metallic nano-clusters which could 
also be responsible for this observed ferromagnetism \cite{KimD}. 
First principles calculations for Co substituting Ti sites in rutile as well as anatase TiO$_{2}$
show creation of Co $3d$ bands at the Fermi energy implying metallicity \cite{Geng, Weng2}. 
Also  a net magnetic moment of ${\sim}$0.6${\mu}_{B}$ occurs at the Co site suggesting that Co 
is in low spin state.

Implantation is a less explored method for introducing magnetism in a non-magnetic system. By this technique,
small concentrations of dopants can be introduced, at well defined depth, in the host. Systems produced
in this fashion can show many interesting electronic and magnetic properties \cite{Xin}. Mn implanted GaP displays  
enhanced magnetism with a transition temperature above 300~K \cite{Overberg}. Mixed phases of 
ferromagnetism and super-paramagnetism have been observed after high fluence (1.5x10$^{17}$~ions/cm$^{2}$)
implantation of Cobalt in rutile TiO$_{2}$(001) single crystals \cite{Akdogan, Khaibullin} which forms the basis of present study.
Few studies have also shown SPM behavior in Co thin films on TiO$_2$ \cite{Shinde}.  
However, none of these studies have investigated  the anisotropic nature of magnetization along different 
crystallographic directions, or the role of inter-cluster interactions. 

The present study investigates the magnetic behavior after Cobalt implantation in rutile TiO$_2$(110)
single crystals. Surprisingly, a super-paramagnetic behavior is observed here instead of the expected 
ferromagnetic behavior from TiO$_2$ which is a DMS material. 
Here a detailed study of magnetic behavior, originating especially from super-paramagnetism, both above 
and below the blocking temperature, is presented. Interestingly, the super-paramagnetic behavior here 
is anisotropic in nature, i.e. non-equivalent along the two crystallographic axes of TiO$_{2}$. Development of 
Ti$_{1-x}$Co$_{x}$O$_{2}$ phase and Cobalt nano-clusters give rise to properties not observed before 
in SPM or DMS based systems.

\section{Experimental Details}
\label{exp}
Commercially available single crystals ($5mm\times5mm\times1mm$) of TiO$_{2}$ (from Matek)
with $<$110$>$ crystallographic direction perpendicular to the surface, were implanted with
cobalt ions at room temperature with fluences of 5$\times$10$^{16}$,  8$\times$10$^{16}$ 
and 1$\times$10$^{17}$~ions/cm$^{2}$. These samples have been labeled here as A, B and C,
respectively. In addition, implantation was also carried out at two lower fluences of 
1$\times$10$^{16}$ and 3$\times$10$^{16}$~ions/cm$^{2}$.  Co ions were implanted in 
TiO$_2$ with an energy of 200~keV. The penetration depth of Co in TiO$_2$ has been
evaluated using SRIM to be $\sim 90nm$ \cite{Ziegler}. 
The structural modifications have been investigated using X-Ray Diffraction (XRD), both in
conventional $\theta-2\theta$ geometry as well as in grazing angle geometry, on a 
Bruker diffractometer, using Cu K$_{\alpha}$ source. For grazing incidence XRD studies,
an incidence angle of 2$^0$ was chosen. Magnetic measurements were performed 
using a commercial Superconducting Quantum Interference Device (SQUID). Temperature 
dependence of magnetization ($M$) has been obtained for Zero-Field Cooled (ZFC) 
as well as Field Cooled (FC) conditions in a field of 0.05~T. Magnetization ($M$) vs.
magnetic field ($H$) measurements have been carried out at various temperatures ranging 
between 2 and 300~K. The magnetic measurements have been carried out with H parallel ($H_\parallel$)
as well as perpendicular ($H_\perp$) to $<$001$>$ crystallographic direction of 
the TiO$_{2}$ crystals. 

\section{Results and discussion}
\label{result}
\subsection{X-Ray Diffraction}
\begin{figure}
\centering
\includegraphics[width=0.60\textwidth]{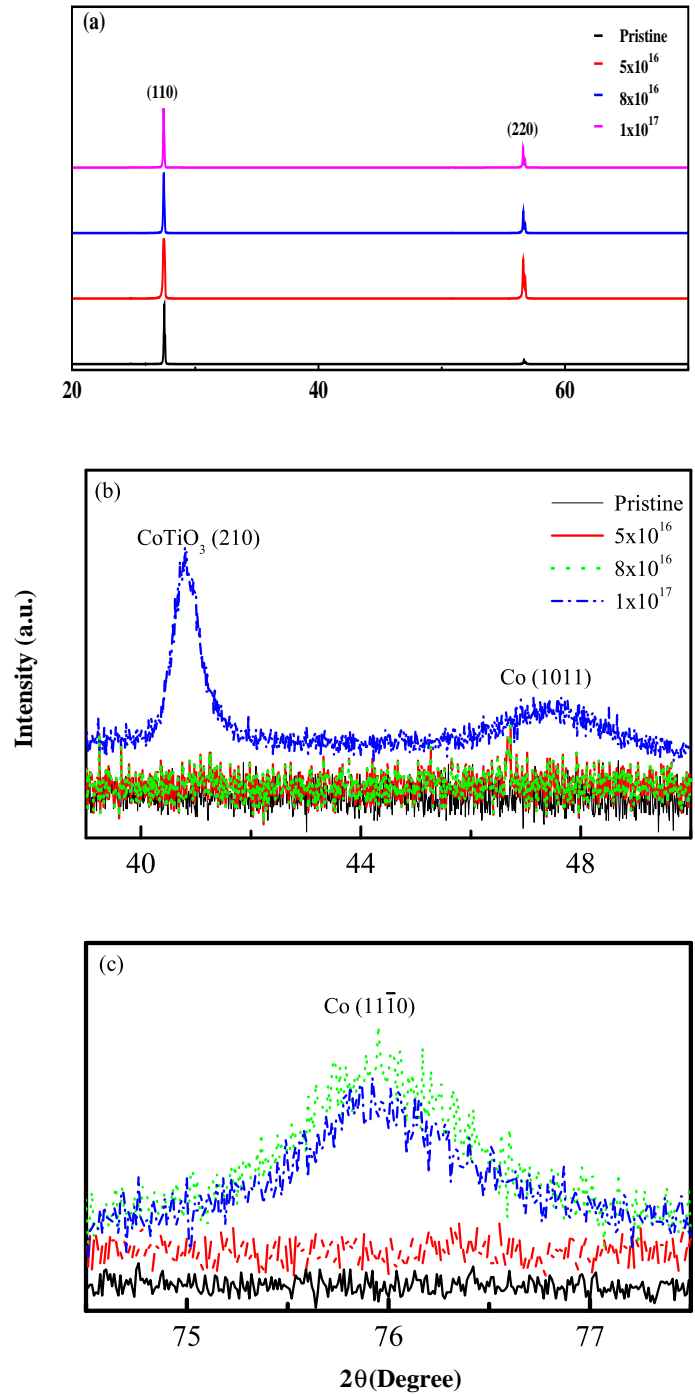}
\caption{ (a) XRD in the normal ${\theta}-2{\theta}$ geometry shows (a) [110] and [220]
planes of TiO$_2$(110) for the pristine and after Cobalt ion implantation (b)
formation of Co clusters and secondary CoTiO$_{3}$ phase at the fluence of 
1$\times$10$^{17}$~ions/cm$^{2}$ (c) Grazing incidence XRD showing Co clusters for 
8$\times$10$^{16}$ and 1$\times$10$^{17}$~ions/cm$^{2}$.  }
\label{fig:fig1}
\end{figure}

Fig.~\ref{fig:fig1} displays the XRD results from pristine TiO$_2$ as well as after it is implanted 
with Co at the fluences of  5$\times$10$^{16}$, 8$\times$10$^{16}$ and 
1$\times$10$^{17}$~ions/cm$^{2}$ (samples A, B and C). Fig.~\ref{fig:fig1}(a)
shows a sharp feature at 27$^{o}$ both prior to and after implantation. Fig.~\ref{fig:fig1}(a)
shows the $<$110$>$ Bragg reflection at 27$^{o}$ in the pristine and implanted TiO$_{2}$ crystals. 
The Bragg peaks show slight shift and become little broader upon Cobalt implantation indicating 
generation of some stress as well as substitutional incorporation of Co in TiO$_2$ lattice 
with the formation of Ti$_{1-x}$Co$_{x}$O$_{2}$ (x $<$ 0.01) at low fluences \cite{Shalik}. 
Fig.~\ref{fig:fig1}(b) presents the normal XRD (39-50$^o$) results which show presence of Cobalt clusters
at the highest fluence of 1$\times$10$^{17}$~ions/cm$^{2}$, not observed for lower fluences. Here,
the broad feature at ${\sim}$ 47.4$^{o}$ reflects the formation of hexagonal closed packed (hcp) Cobalt (1011) clusters.
In addition, the feature at ${\sim}$ 40.7$^{o}$ suggests formation of a secondary CoTiO$_{3}$ (210) phase 
\cite{Newnham}. Development of  Cobalt clusters is also reflected by the grazing incidence 
XRD studies (fig.~\ref{fig:fig1}(c)) where hcp Co(11$\bar{1}$0) is observed at 76$^{o}$ for the samples~B and C but not for the sample~A. Based on the 
width of hcp-Co Bragg peak, particle size of hcp-Co cluster has been determined to be
${\sim}$7~nm. The effect of cluster formation on the magnetic properties is discussed below.

\subsection{Magnetization vs. Temperature}
\begin{figure}
\includegraphics[width=1.0\textwidth]{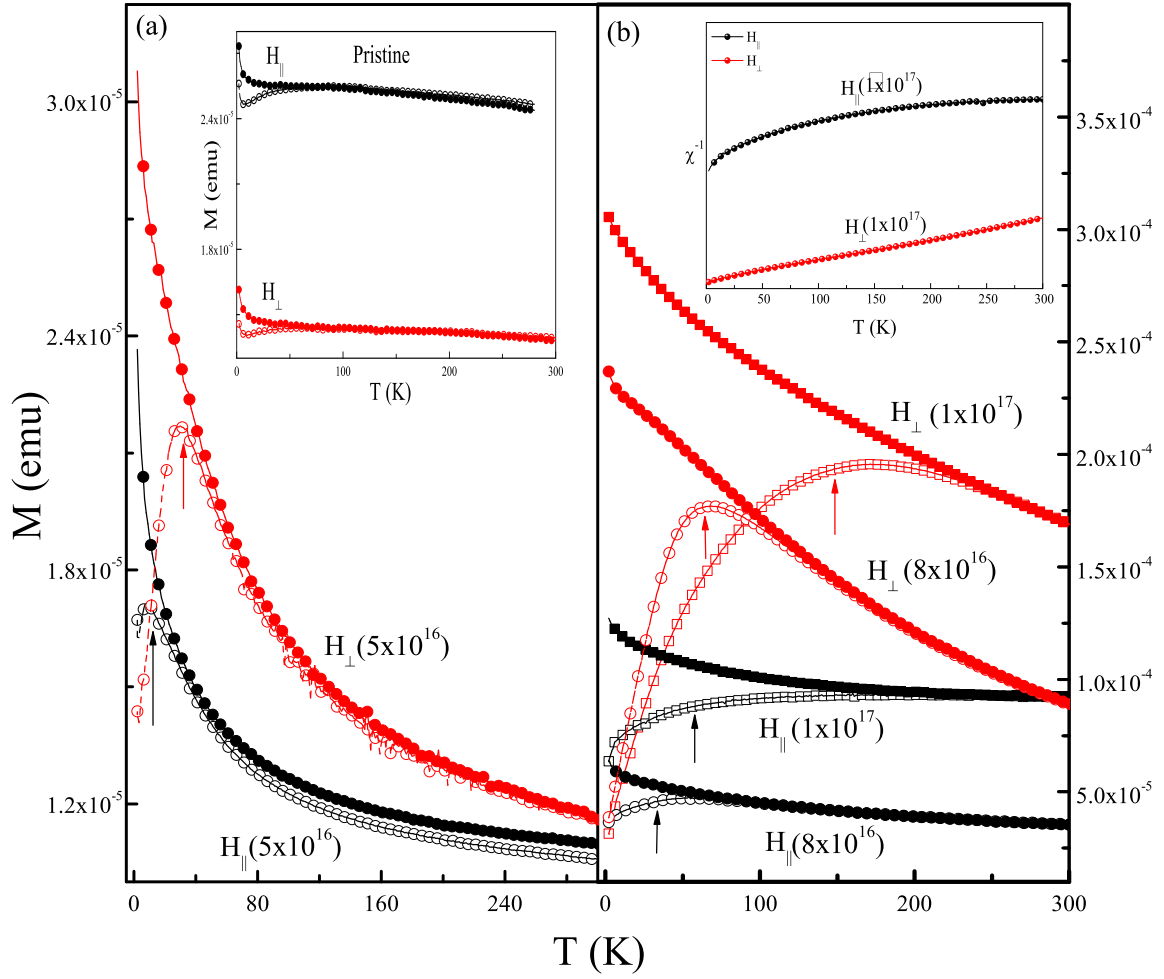}
\caption{ ZFC (open symbols) and FC (closed symbols) graphs for cobalt implanted TiO$_{2}$ at 500~Oe for (a) sample~A,
 inset shows for pristine. (b) samples~B and C, inset shows $\chi^{-1}$
for sample C.
The black and red symbols correspond to field parallel ($H_\parallel$) and perpendicular
($H_\perp$) to $<$001$>$ directions  of TiO$_{2}$ crystal. The arrows indicate the 
blocking temperatures.}
\label{fig:fig2}
\end{figure}

The temperature variation of magnetization for both pristine as well as implanted samples were measured 
with magnetic field (H) pointed along two crystallographic directions of TiO$_{2}$ viz. $<$001$>$ 
(H$_\parallel$) and $<$1${\bar 10}$$>$ (H$_\perp$). Fig.~\ref{fig:fig2}(a) (inset) displays the ZFC-FC plots
for the pristine sample. Though the trends in magnetization are the same, along both the field
directions, magnetic moment (M) is higher along H$_\parallel$. This suggests H$_\parallel$ 
($<$001$>$) to be the easy (anisotropic)-axis in pristine TiO$_{2}$. Below 20~K, a downward trend in the ZFC plot is observed
which rises again upon further decrease in temperature. Ideally, the pristine sample should be 
non-magnetic due to the empty $d$ orbital of Ti$^{4+}$. However, Van Vleck paramagnetism and a 
weak defect induced magnetic ordering is observed due to the presence of O and Ti type vacancies 
and other defects \cite{Park, Errico}. 

For small Co implantation fluences (1$\times$10$^{16}$ and 3$\times$10$^{16}$ ions/cm$^{2}$),
though there is a slight increase in the magnetization (data not shown ), the nature of 
ZFC-FC curves remain similar to the pristine. At these low fluences, lattice remains paramagnetic
with Co ions substitutionally incorporated in the TiO$_2$ lattice and net magnetic behavior
unaltered compared to pristine. However, the magnetic moment is higher for H$_\perp$ indicating
this to be the preferred magnetization direction with Co substitution.

For the Co fluence of 5$\times$10$^{16}$ ions/cm$^{2}$ (sample A), a drastically different
magnetic nature compared to the pristine is observed. Fig.~\ref{fig:fig2}(a) shows the ZFC-FC plots for both
the field directions. For H$_\parallel$, a bifurcation in ZFC is noticed
at 8~K, while for H$_\perp$ this is at 30~K. With increasing fluence, i.e. 
for sample~B and C, the bifurcations shift to much higher temperatures. Fig.~\ref{fig:fig2}(b) shows the 
ZFC-FC plots for samples B and C in both field configurations. The nature of splittings 
in these plots indicate that samples A, B and C show super-paramagnetism, with arrows indicating
the respective blocking temperatures (T$_{B}$). The large increase in T$_{B}$ with fluence 
indicates an increase in the particle size while the broadening of the transition around 
T$_{B}$ indicates a large variance in particle diameters. For all the three samples, 
$M$ and $T_{B}$ are considerably higher for field along H$_\perp$ compared to
H$_\parallel$ (also see table~1). At the highest fluence 1$\times$10$^{17}$~ions/cm$^{2}$ 
(sample C), the difference in T$_{B}$ of 80~K between $H_\parallel$ and $H_\perp$,
indicates that this system is very anisotropic. 

For H$_\parallel$ field, magnetization in samples B and C appears nearly constant, 
with temperature, above  $T_{B}$. The decrease in magnetization is much slower than 
the expected Curie-Weiss like decrease in SPM systems. However for H$_\perp$ field, 
a decrease in magnetization with temperature is observed simalr to that expected in SPM systems. 
Fig.~\ref{fig:fig2}(b) (inset) shows a plot of inverse susceptibility ($\chi^{-1}$)
as a function of temperature, for sample~C, along both the field directions. Similar to a paramagnetic system, 
$\chi^{-1}$ should show a linear increase with temperature. However, here
this is only observed in the case of H$_\perp$ direction. The intercept on the T-axis is negative, 
indicating presence of some antiferromagnetic couplings in the system, similar to the
pristine. For H$_\parallel$, $\chi^{-1}$ is almost constant above $T_{B}$ (fig.~\ref{fig:fig2}(b) inset). 

Thus the magnetic moments of the 
clusters in this cobalt implanted system are rotatable, due to external field and temperature, only
when the field is along $H_\perp$ direction. 
In a typical SPM, well below T$_{B}$, the ZFC susceptibility increases with increasing 
temperature, suggesting that more nano-particles get unblocked and contribute to the 
susceptibility. Most of the particles get unblocked near $T_{B}$ and the system becomes 
an SPM for $T > T_{B}$. Thus irrespective of field
direction all SPM particles get unblocked above $T_{B}$ and display a Curie-Weiss like
behavior with increasing temperature. Such a behavior 
is observed, above $T_{B}$, in our system only for 
$H_\perp$. For field along $H_\parallel$, the magnetization above T$_{B}$ remains 
nearly constant with increasing temperature. This suggests that along this direction, 
only the smaller nano-particles get unblocked above $T_B$, whereas the larger particles 
remain blocked. Combined anisotropy due to the Co clusters and Ti$_{1-x}$Co$_{x}$O$_{2}$
will be responsible for this observation, as will be discussed below.

\subsection{Magnetization vs. Field}
\subsubsection{Above Blocking temperature}
\begin{figure}
\centering
\includegraphics[width=0.85\textwidth]{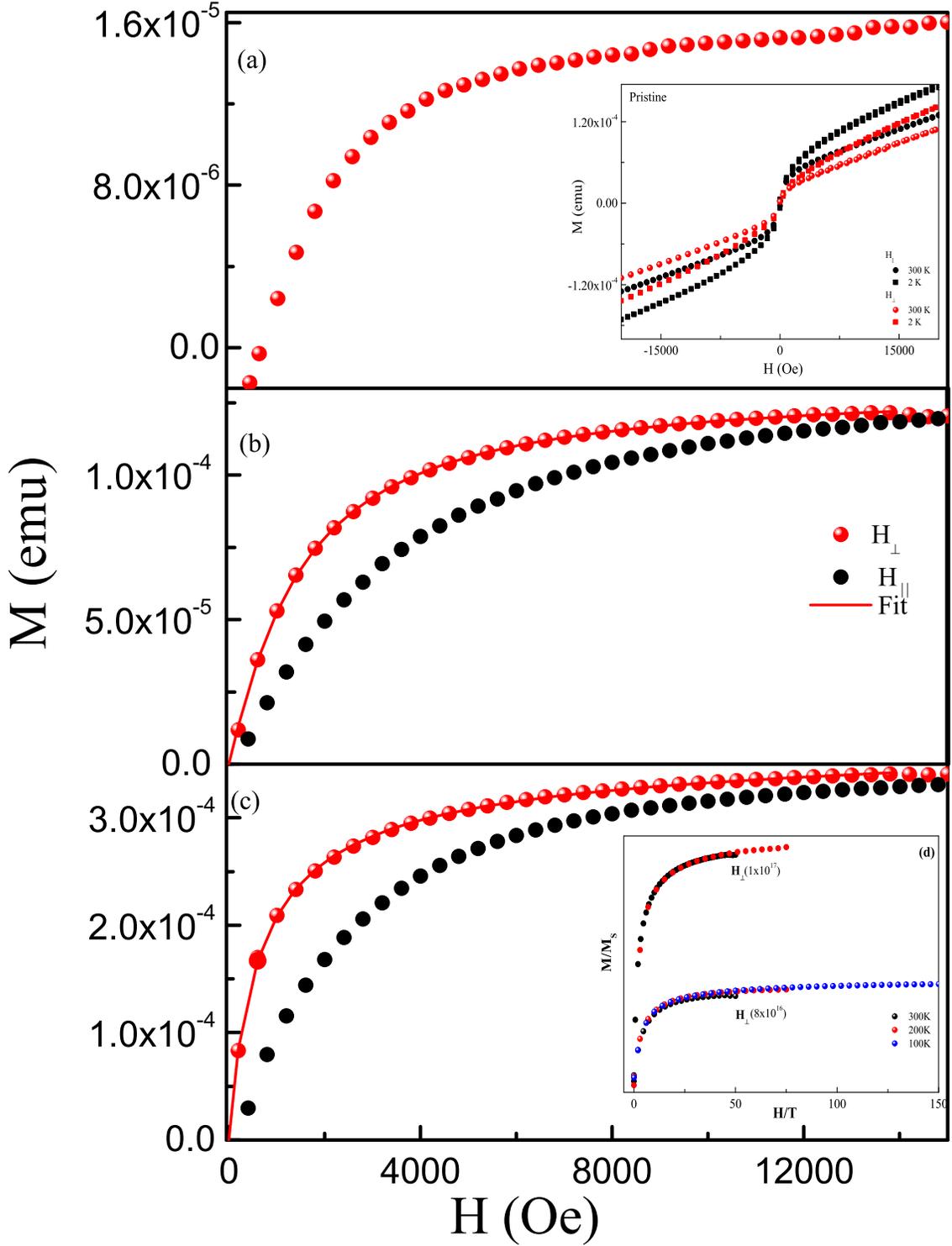}
\caption{ Magnetization v/s Field graphs for Cobalt implanted TiO$_{2}$ 
for field along $H_\perp$ (red symbols) and $H_\parallel$ (black symbols).
The solid curves correspond to fitting to Langevin function with lognormal 
distribution (eqn~2). In (a) data of sample A corresponds to T = 100~K. 
Inset shows the magnetization vs field of pristine sample obtained at T = 300 and 2~K, 
for both field directions. (b) Experimental and calculated (using eqn.~2)
$M$-$H$ plots at T = 300~K for sample~B along both field directions. 
(c) Experimental and calculated (using eqn.~2) $M$-$H$ plots at
T = 300~K for sample~C along both field directions. (d) Universal scaling behavior
of samples~B and C for field along $H_\perp$ at various temperatures.}
\label{fig:fig3}
\end{figure}

Fig.~\ref{fig:fig3} shows the magnetization isotherms, above $T_B$, for pristine and cobalt implanted samples. 
$M$-$H$ plots for the pristine TiO$_{2}$, at 2 and 300~K, are shown in the inset (of fig.~\ref{fig:fig3}(a)) 
for $H_\parallel$ and $H_\perp$. At 300~K, the magnetization 
rises linearly as expected for a paramagnetic TiO$_2$, without attaining any saturation. 
A small ($<$ 20 Oe) coercivity has also been observed.
Moreover, the magnetization is higher along $H_\parallel$ (than $H_\perp$) indicating
 this to be the easy axis of magnetization in the pristine sample.  Similar behavior 
is also observed at 2~K. The magnetization curve for 
Sample~A at 100~K along $H_\perp$ is shown in fig.~\ref{fig:fig3}(a) and shows a saturation- like behavior near
1~T. This curve also displays a very small coercivity of 30~Oe.
At this fluence the cobalt concentration in TiO$_2$ lattice is small and consequently the 
Co induced magnetic moment is comparable to the pristine. Assuming that the magnetization from the host lattice and from implanted ions 
are independent in any implanted system, so the the contribution 
to magnetization of the former has been subtracted from that of all the implanted samples. After subtracton of the pristine magnetization,
the magnetization plots show a saturation like behaviour for all temperatures, which is the true behaviour of the SPM system.

The magnetization plots for samples~B and C, at 300~K, are shown in fig.~\ref{fig:fig3} for
$H_\parallel$ and $H_\perp$. Both the samples show higher magnetization when the 
field is along $H_\perp$ (than $H_\parallel$). While the easy-axis in pristine is along $H_\parallel$, these (B and C) show higher magnetization for $H_\perp$ which
indicates a reversal in the preferred direction
of anisotropy. Moreover for both the samples~B and C, the slope of magnetization at low H, is much steeper
for fields along $H_\perp$ than $H_\parallel$. As a result, the near-saturation like behavior
is attained faster (at field $\sim$~0.2~T) in the former case than in the latter case where
saturation is achieved at fields around 1~T. Ideally for an SPM system, the $M$-$H$ isotherms 
should be reversible \cite{Bean}. However both samples B and C, show a mild irreversible 
behavior. For $H_\perp$, small coercivities of nearly 60 and 90~Oe are observed
for samples~B and C. Coercivities are smaller along $H_\parallel$.

For an SPM system, a plot of $M$/$M_{S}$ vs. $H/T$ ($M_{S}$ is saturation magnetization) 
should scale into a single universal curve \cite{Luis}. Here, no scaling has been observed 
for sample~A. Samples~B and C also do not show any scaling for field along $H_\parallel$. However 
for $H_\perp$ field, interestingly, scaling behavior is displayed (see fig.~\ref{fig:fig3}(d)). The scaled
curves for sample~B show that though the scaling exists, there are some deviations. A possible
reason can be the existence of long ranged antiferromagnetic couplings due to the formation of
Ti$_{1-x}$Co$_{x}$O$_{2}$, that results in an effective molecular field which hinders
complete SPM-like behavior at this stage. For samples~ C, a nearly perfect scaling is observed indicating a 
good SPM character. Absence of scaling along $H_\parallel$ can be due to the anisotropic effects 
that restrict free rotation of magnetic moments, of the nano-particles, for field applied along this direction.
Thus similar to ZFC-FC plots of fig.~\ref{fig:fig2}, the magnetization results of fig.~\ref{fig:fig3} show that the
SPM -like behavior is observed in samples B and C only, when the field is applied along $H_\perp$.

In a superparamagnet, there exists a distribution of magnetic moments due to the variations
in the particle size of the nanoclusters. Hence, the net magnetization is given as a weighed
sum of the Langevin function \cite{Chantrell},
\begin{equation}
\label{eqn2}
M(H, T)= \int_{0}^{\infty} \mu L\left(\frac{\mu H}{k_{B}T}\right) f(\mu) d\mu
\end{equation}
\noindent where $L(x)$ = $coth(x)-1/x$ is the Langevin function, $f$(${\mu}$) is 
the distribution of magnetic moments, given by a log-normal distribution \cite{Fonseca},
\begin{equation}
\label{eqn3}
 f(\mu)= \frac{1}{\sqrt{2\pi}\mu\sigma}exp\left[-\frac{ln^{2}(\frac{\mu}{\mu_{0}})}{2\sigma^{2}}\right]
\end{equation}
Here ${\mu}_{0}$ is the median of distribution and ${\sigma}$ is the width
of this distribution. The mean magnetic moment ${\mu}_{M}$ = ${\mu}_{0}$exp($-{\sigma}^{2}$/2).
Assuming all the nano-particles to be spherical, ${\mu}_{0}$= ${\pi}M_{S}D^{3}$/6.
Here, D is the diameter of particles and $M_{S} (= 1.56~u_{B}$) is the saturation magnetization 
of bulk cobalt. The above expression holds true for T$>>$T$_{B}$, where the role of anisotopy can be neglected. 

In many SPM systems, equation (2) has been shown to yield good results for magnetization
well above $T_B$. The fittings for the magnetization curves, using this eqn.~(2), have also
been shown in fig.~\ref{fig:fig3}(b,c) for samples B and C when the fields is along $H_\perp$.
Fittings of the magnetization curves have been utilized to obtain average magnetic 
moment (${\mu}_{M}$), particle size (D) and standard deviation ($\sigma$) and are listed in table~1.
With increasing fluence there is a systematic increase in the average magnetic moment as well as the
particle diameter. In addition, the deviation ${\sigma}$ also considerably 
increases from sample A to C. This is also observed in the ZFC-FC plots, wherein 
sample A displays a sharper transition while sample C shows a broader transition.

\subsubsection{Below blocking temperature} 
\begin{figure}
\includegraphics[width=1.0\textwidth]{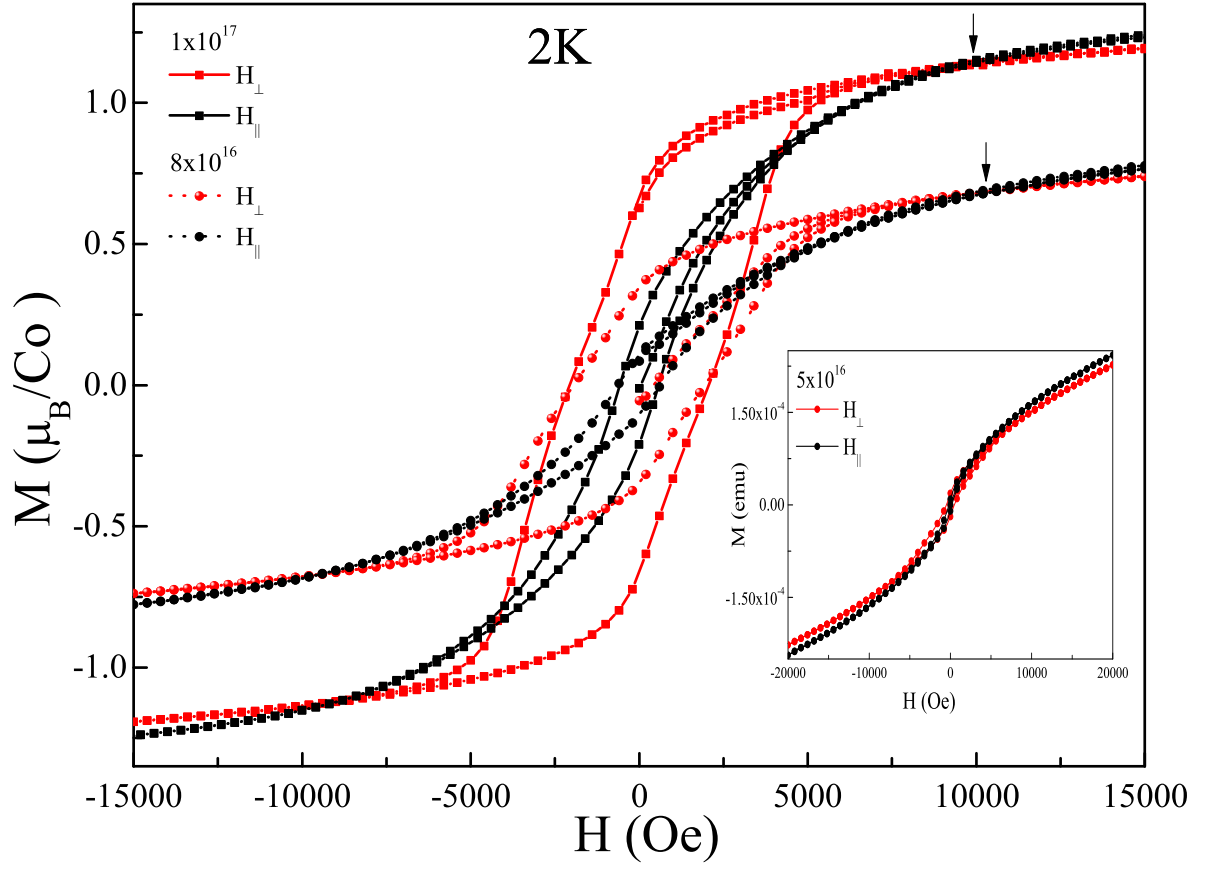}
\caption{ $M$-$H$ curves of samples B and C obtained at 2~K for fields along 
$H_\perp$ (red symbols) and $H_\parallel$ (black symbols). Arrows indicate 
cross-over. The inset shows $M$-$H$ data for sample~A at 2~K. }
\label{fig:fig4}
\end{figure}

The magnetization curves for the implanted samples below T$_B$, at 2~K, are
presented in fig.~\ref{fig:fig4}. Inset shows the magnetization curves of sample~A for both 
$H_\perp$ and $H_\parallel$. Small coercivity is observed along both 
these fields. Magnetization curves for the pristine at 2~K are similar to those of  
sample~A. Also as demonstrated by sample~A (fig.~\ref{fig:fig4} inset), 
the magnetization in pristine does not saturate even at high fields (2~T) but rather continues to 
increase, indicating a paramagnet- like behavior. 
  
Fig.~\ref{fig:fig4} shows the hysteresis behavior of samples B and C along $H_\parallel$ and $H_\perp$.
For samples~B and C, coercivities ($H_{C}$) as large as 1500 and 1800~Oe, respectively,  
are observed for $H_\perp$ field. These coercive fields are nearly 3 times 
larger than those observed for Fe implanted TiO$_2$ \cite{Shengqiang} 
or for nano-Cobalt systems prepared by other methods \cite{Luo}.

For both the samples~B and C, the magnetization ($M-H$) plots display a near saturation-like behavior
above 1~T (fig.~\ref{fig:fig4}). Moreover, the slope ($dM/dH$) here at 2~K is higher than that observed above 
T$_B$ at 300~K (fig.~\ref{fig:fig3}(b,c)). A crossover in $M$ between the two field directions (indicated 
by the arrows in fig.~\ref{fig:fig4}) is also observed at 2~K for both the samples. This suggests 
an additional contribution to the magnetization at low temperatures and high fields which
will be discussed below in the section on {\it anisotropic paramagnetism}. 
For 1.5~T field, both the samples~B and C show a near-saturation
like behavior with $M_{S}$ ${\sim}$ 0.8 and 1.2 ${\mu}_{B}$/Co~atom, respectively.
These values are considerably lower than the saturation magnetization (1.56 ${\mu}_{B}$/Co~atom)
for bulk cobalt \cite{Akdogan}. Since the presence of isolated Co-atoms or sub-nano few-atom Co clusters will
effectively not contribute to the magnetization, this observation suggests presence of some
Co atoms or smaller clusters in the TiO$_2$ lattice. Only the nano-dimensional
or bigger Co clusters give rise to the observed magnetization and estimates here show that for samples
A, B and C nearly 22, 43 and 67$\%$ of the implanted cobalt atoms, respectively, form such clusters.
    
Starting from 2~K, the width of hysteresis loop decreases on increasing the temperature, in 
both the field directions for samples B and C. $M-H$ loop for both the samples display trends like 
a hard ferromagnet, similar to metallic Fe and Co. The 
hysteresis loop appears like a ``parallelogram" for $H_\perp$, indicative 
of the easy axis of magnetization, but narrow ``ribbon-like" for
$H_\parallel$, similar to the hard axis of a ferromagnet.
	
\section{Discussion}
\subsection{Anisotropy}
\subsubsection{Super-paramagnetic region}
We briefly discuss the effect of anisotropy on the magnetic behaviour of Co nano clusters in the super-paramagnetic regime. We demonstrate that the Co nano-particles
are not randomly oriented, like usual SPM clusters, rather have a fixed easy axis direction even above the blocking temperature.
When $T > > T_{B}$, effects of uniaxial anisotropy can be neglected and magnetization can be completely described 
by eqn.~(2). However as temperature reduces, effects of anisotropy become non-negligible and 
the magnetization cannot be described in a simple analytical manner 
as in eqn.~(2). For a system of SPM clusters, each with a random anisotropy direction, anisotropy $K$, saturation magnetization $M_{S}$  
and fixed volume $V$, the magnetization is obtained from the Hamiltonian, ${\mathcal{H}}= -V({M}_{S}{H} cos(\alpha) +Kcos^{2}({\theta}))$.  
Here, the first term corresponds to the external magnetic field energy and the second term corresponds to easy axis anisotropy energy. 
The external magnetic field $H$ makes an
angle ${\alpha}$ with the magnetization and ${\lambda}$ with the easy axis, while ${\theta}$ corresponds to the angle between the easy axis and the magnetization.
The full configuration is described in detail by Morup {\it et al.} \cite{Morup}, in which the net magnetization $M$($H$, $T$, $K$) is obtained after
integration over the angles ${\theta}$, ${\alpha}$, the azimuthal angle ${\phi}$ and finally over all possible directions of applied field.
However, in our case we consider two extreme cases viz. ${\lambda}$ = 0 and ${\pi}$/2, corresponding to the easy axis ($H_\perp$) and the 
hard axis ($H_\parallel$) respectively.
 
Using this formalism, the magnetization curves for the two ${\lambda}$ values have been calculated for T = 100 K, $M_{S}$ = 1.2 ${\mu}_{B}$,
$K$ = 5x10$^{5}$J/m$^{3}$, $V$ = 41x10$^{-27}$m${^3}$ which corresponds to the sample with highest fluence, and are shown in fig. 5.
The nature of theoretical curves, for two field directions, agree well the experimental isothermal
magnetization results shown in fig.~4. Thus the theoretical results clearly show that applying the fields
along $<$1${\bar 10}$$>$ and $<$001$>$ crystallographic directions is exactly equivalent to applying magnetic fields along the ``easy" and ``hard" axis of the Co
clusters. 

This confirms that the magnetic anisotropy directions of the clusters, in the present system, are not entirely random
as expected in a super-paramagnetic system but are effectively along the $<$1${\bar 10}$$>$ crystallographic direction.
This also suggests that the anisotropy of the Co clusters is coupled to the magneto-crystalline
anisotropy of the TiO$_{2}$/Ti$_{1-x}$Co$_x$O$_{2}$ lattice which would be subsequently discussed.

\subsubsection{Anisotropy Below the Blocking Temperature}
The $M-H$ plots of samples~ B and C for 2 K show a large anisotropic character similar to that of the bulk metallic Co.
The uniaxial anisotropy of the system is given by $K_{\mu} = M_{S}H_{K}/2$, where $H_{K}$ is 
the anisotropy field \cite{Shengqiang}, which is also the saturation field for the hard axis. In both the samples,
the saturation along the hard axis ($H_\parallel$) is
attained at $H_{K} \sim~1.2$~T, suggesting anisotropy constant of 2.5 $\times$ 10$^{5}$~Joule/m$^{3}$ 
and 5.39 $\times$ 10$^{5}$~Joule/m$^{3}$ for samples B and C, respectively.
These values are smaller than the anisotropy constant of 7.5 $\times$ 10$^{5}$~Joule/m$^{3}$ observed for the bulk 
cobalt at 5 K \cite{Wiekhorst}. This is due to the presence of some isolated (substituted) atoms and sub-nano clusters
in the TiO$_2$ lattice which also led to (see fig.~4) lower saturation magnetization for samples~B and C, compared to
the bulk Cobalt. Bulk cobalt has a uniaxial anisotropy along the hexagonal $c$ direction which 
corresponds to the easy axis \cite{Kim}. Thus in samples B and C, it can be expected that the $c$ axis of the individual
cobalt clusters should be aligned along the (1${\bar1}$0) direction of TiO$_{2}$ crystal. Similar oriented metallic
clusters have also been observed in Fe implanted TiO$_{2}$ where $<$1${\bar 1}$0$>$ 
direction ($H_\perp$) of TiO$_{2}$ is the easy axis \cite{Shengqiang}. However, does not show any specific crystallographic
orientation of the Co clusters. This could be because, unlike cubic symmetry of Fe, the hexagonal symmetry of Co is not
compatible with the tetragonal TiO$_{2}$ crystal (see fig.~1). Hence, the origin of the strong anisotropic character of the SPM clusters could be due to the combined effect of
the uniaxial anisotropy of the hexagonal Co clusters along with the magneto-crystalline anisotropy of the host TiO$_{2}$/Ti$_{1-x}$Co$_{x}$O$_{2}$ lattice.

In switching of the easy axis, from being along $H_\parallel$ ($<$001$>$) prior to implantation 
to $H_\perp$ ($<$1${\bar 1}$0$>$) after implantation, it is assumed that the anisotropy of Co spins
occupying Ti sites play a significant role. To verify this, first principle calculations
have been carried out to determine the magneto-crystalline anisotropy energy of 
Ti$_{1-x}$Co$_{x}$O$_{2}$ using VASP \cite{Kresse}. The calculations were performed for 
three spin directions of Co spin: along $H_\parallel$, along $H_\perp$ and along
 $<$110$>$. The net magnetic moment $\sim$0.7$~{\mu}_{B}$ obtained here is lower than 
1~${\mu}_{B}$ expected for the $S$=1/2 system, suggesting an itinerant character. Among the three 
spin configurations, the system has the lowest energy when spin is along $H_\perp$, while
energy is highest along $H_\parallel$ direction. This is in agreement with the experimental
results observed here. The difference in energy, in these two directions, is approximately -0.5~meV/Co i.e.
nearly $\sim$5~K. These results indicate that the magnetocrystalline anisotropy of Co in 
Ti$_{1-x}$Co$_{x}$O$_{2}$ determines the easy and hard axis and leads to highly anisotropic 
super-paramagnetism in TiO$_{2}$.

\subsubsection{Anisotropic paramagnetism}
In addition to the anisotropic effects of super-paramagnetic Co clusters, an additional
anisotropy from paramagnetic Co ions has also been observed here. In the magnetization 
isotherms of samples ~B and C above $T_B$ (fig.~\ref{fig:fig3}), magnetization is observed to be nearly
constant for fields higher than 1~T, especially for field along $H_\perp$. However in the blocked region,
at 2~K, an increase in $dM/dH$ for samples~B and C is observed (see fig.~\ref{fig:fig4}). This is in 
contrast to the usual SPM systems where $M$ remains almost constant with increasing H.
A possible reason for this could be the presence of uncompensated paramagnetic Co spins. 
In addition, a cross over in magnetization between the two field directions (shown by
arrow in fig.~\ref{fig:fig4}) is also observed. The cobalt that occupies Ti sites in TiO$_{2}$ lattice
should be in a 4$^+$ state corresponding to the $S$=1/2 system, as also observed in
the first principles calculations carried out here. In $S$=1/2 system, the single ion 
anisotropy does not affect the magnetic behavior. The unusual cross-over behavior can be 
explained qualitatively by considering the presence of Co$^{2+}$ ions, i.e. $S$=3/2 system. 
The crossover observed here then arises due to presence of the single ion anisotropy of this 
3/2 spin state. Origin of Co$^{2+}$ is via formation of CoTiO$_{3}$ nanoclusters whose
presence in sample~C has been observed by XRD (fig.~\ref{fig:fig1}). In addition, Co$^{2+}$ also exists as 
Ti$_{2}$O$_{3}$. XPS studies have shown that along with a Ti$^{4+}$ 
state, a small percentage of Ti$^{3+}$ also develops for samples B and C and increases with the fluence 
\cite{Shalik}. Still, the main source of Co$^{2+}$ spins here is CoTiO$_{3}$ and 
though it has been observed only for sample~C, it is likely that smaller amounts will be
present in sample~B also. 

Similar paramagnetic anisotropy has been observed in the magnetization of Co:ZnO thin films 
\cite{Ney}. Ney et al. \cite{Ney, Ney1} have discussed the magnetization in terms
of an {\it effective spin Hamiltonian} with an anisotropy along the crystallographic $c$ axis 
(referred as $z$ axis) of Zn$_{1-x}$Co$_{x}$O films. Here, 
effective spin Hamiltonian has been applied to understand the anisotropic paramagnetic behavior
observed in fig.~\ref{fig:fig4}. Remarkably, no crossover was observed when anisotropy was along $\hat{z}$ axis (i.e.
along $H_\parallel$: $<$001$>$), rather it was observed when $\hat{x}$ direction (along 
$H_\perp$: $<$1${\bar 1}$0$>$) of the spin was considered to be the axis of single ion 
anisotropy (geometry of the present system is shown in fig.~5). Hence, a modified $S$ = 3/2 spin Hamiltonian 
was used here and is discussed below:
\begin{equation}
\hat{H}_{spin} =  {\mu}_{B}g_{\parallel}H_{z}S_{z}+ {\mu}_{B}g_{\perp}(H_{x}S_{x}+H_{y}S_{y})+QS_{x}^{2} ,              
\end{equation}
Here the magnetic state is characterized by two g factors,
g$_{\parallel}$ (2.238) and g$_{\perp}$ (2.276), and the zero field splitting 
constant $Q$ \cite{Ney}. In the above equation, $Q$ corresponds to single ion anisotropy of $S$=3/2
system, due to Co$^{2+}$, along the $\hat{x}$ direction. Spin Hamiltonian has been applied 
to calculate the energy levels of the $S=3/2$ system $|M_{S}>= |-3/2> \cdots |+3/2>$ by using the
matrix $<M_{S}|\hat{H}_{spin}|M_{S}>$ for H $\parallel$ $\hat{z}$ (H $\parallel$ $H_\parallel$) and
H $\parallel$ $\hat{x}$ (H $\parallel$ $H_\perp$). Diagonalization of the matrix provides four eigenvalues
along each direction. These energy values E$_{i,a}$ ( $i : M_{S}$ values and $a = H || \hat{z},\hat{x}$)
were used to calculate the magnetization $M = ({\partial}F/{\partial}T)_{H}$ of the magnetic free 
energy $F=-k_{B}Tln(Z_{i,a})$ using the partition function $Z_{i,a} = \sum_{i} e^{-E_{i,a}/k_{B}T}$. 

Fig.~\ref{fig:fig6} shows the $M-H$ curves calculated at T= 2~K using the spin 
Hamiltonian for varying strengths of Q. The strength of zero-field splitting Q was varied from 0~K 
to 4~K and its role on the anisotropy of the $M-H$ curves was investigated. For $Q=0$~K the $M - H$
curves calculated for H $\parallel$ $H_\parallel$ and H $\parallel$ $H_\perp$ show no cross-over.
The shape of $M -H$ curves for H $\parallel$ $H_\parallel$ do not show much change upon 
increasing $Q$. On the other hand, the $M - H$ curves for H $\parallel$ $H_\perp$ show a decreasing
slope with increasing $Q$. This becomes responsible for an increase in anisotropy, and a cross-over 
(marked by the arrow) is seen for $Q=4$~K. This plot calculated at $T=2$~K can be compared with the
experimental data presented in fig.~\ref{fig:fig4} and displays a similar anisotropic  behavior. 
Moreover, interestingly the effective Hamiltonian calculations for $S=3/2$ system of Co in TiO$_2$ here
show that the anisotropy is observed when $\hat{x}$ ($H_\perp$) direction of the spin is considered to 
be the axis of single ion anisotropy, unlike the $\hat{z}$ ($H_\parallel$) direction in
Co:ZnO system \cite{Ney}.

\subsection{Dipole and inter-particle exchange interactions}
In the blocked region of a super-paramagnet, the parameters of importance are coercivity $H_{C}$, and the reduced remanence ($m_{R}$ = $M_{R}$/$M_{S}$).
For a system of non-interacting 
particles with random anisotropy axes, the reduced remanence should be $\sim$0.5. The values 
of $M_{R}$ and $H_{C}$ are much smaller for field along $H_\parallel$ axis (fig.~4), since 
it corresponds to the hard axis of the clusters. Hence in this section 
field along $H_\perp$ direction is discussed. With increase in fluence the reduced remanence increases 
along with the coercive field. At 2~K, sample~A shows the lowest $m_{R}$ of 0.2, while samples~B 
and C show values of 0.47 and 0.55, respectively, for field along the $H_\perp$ direction. The 
highly reduced remanence in sample A indicates a dominating presence of antiferromagnetic 
interactions \cite{Weili} in the system.
Similar to the reduced remenance, the temperature dependent coercivity $H_{C} (T)$ for a {\it system of randomly
oriented and non interacting particles} displays a behavior given by \cite{Fonseca},
\begin{equation}
 H_{C} (T) = H_{CO} \left(1- \left({ T \over {T_B^0}}\right)^{1/2}\right). 
\end{equation}
with $H_{CO}$ = $K_{ub}/M_{sb}$, where $K_{ub}$ and $M_{sb}$ are the anistropy and saturation magnetization of bulk cobalt. 
$T_B^0$ is the Blocking temperature at zero field, $T_B^0 = {K_{\mu} <V> \over { k_{B}~ln (\tau_{m}/\tau_{0})}}$,
where $K_{\mu}$ is the uniaxial anisotopy constant, $<V>$ is the average particle volume, $\tau_{0}$ and $\tau_{m}$ are the 
characteristic limiting relaxation time and the measurement time. $K_{B}$ is the Boltzman constant. Fig.~7 shows $H_{c}(T)$ as a
function of reduced temperature $k_{B}T/K_{u}$$<$V$>$ for samples B and C along the $H_\perp$. Here 
$<V>$ corresponds to the average volume and has been evaluated using the size(D) of the 
clusters as mentioned in table~1. In order to directly observe the linear relation, 
inset of fig.~7 also displays a plot of $H_{C}$ against $T^{1/2}$ for samples 
B and C. For sample~B, a linear relation has been observed up to T ${\sim}$ 50~K. However 
above this temperature, deviations occur as $T$ approaches $T_{B}$ (=65~K). In sample~C also, 
$H_{C}$ shows deviations beyond ${\sim}$ 50~K, a temperature which is much lower than its 
$T_{B}$ (=150~K). Also, the extrapolation of $H_{C}(T)$ to 0 yields values of 
$T_{B}$ ${\sim}$ 64~K and 100~K for samples~B and C, respectively, 
which are lower than the observed blocking temperatures. The deviation
from linearity suggests a strong presence of inter-particle interactions between the Co clusters

The most prominent interaction among the clusters is the inter-particle dipole interaction.
In addition, there also exists presence of inter-particle exchange interactions.
Using Monte-Carlo (MC) simulations, Kechrakos and Trohidou etal.\cite{Kechrakos1, Kechrakos2} 
have investigated the role of inter-particle dipole and exchange
interactions in determining the coercivity and remanence of the magnetization,
in the blocked region, for a system of SPM nano-particles 
with random uniaxial anisotropic directions. The variations in $m_{R}$ and $H_{C}$ 
have been studied as a function of increasing volume fraction of the magnetic 
nano-particles. Since the size and number of the Co nano-clusters formed inside TiO$_{2}$, increase with fluence, we expect an increase in the volume fraction of
nano-particles. Thus with fluence, the inter-particle dipolar strength increases, which has a demagnetizing effect on the
magnetization, causing a decrease in $m_R$ and $H_C$. This suggests that in addition to the dipole interaction, there exists a strong
inter-particle exchange interaction between the
clusters which can even cause complete ferromagnetic alignment between the clusters. The itinerant nature of the Co occupying Ti sites,
can produce this carrier mediated
exchange interaction between the clusters. Presence of competing dipolar- interactions has also been observed via ZFC-FC curves. 

\section{Conclusion}
\label{conc}
Present study investigates the magnetic properties of single crystals of rutile TiO$_{2}$ after they
are implanted with Co ions. ZFC- FC curves show presence of super-paramagnetic character above $T_B$.
This SPM behavior, seen due to the development of Cobalt nano-clusters in the host lattice, is 
surprisingly anisotropic along the crystallographic directions of the crystal. With this anisotropy,
SPM behavior is observed only along the $<$1${\bar 1}$0$>$ ($H_\perp$) direction which behaves as 
an easy-axis of magnetization, and not along $<$001$>$ ($H_\parallel$). Analysis with Langevin 
function -fitting considers a lognormal distribution of cluster sizes and yields a systematic increase
in magnetic moment as well as particle volume with fluence, above $T_B$. For sample~C, a linear behavior
in inverse susceptibility, higher $T_{B}$ and a good $M/M_{S}$ vs. $H/T$ scaling is observed only when
field is in H$_{\perp}$ direction. Such anisotropy is very unexpected and shows that though
along this $H_{\perp}$ direction magnetic moments are easily rotatable, above $T_B$,  this is not
the case along $H_\parallel$ direction where a considerable fraction of spins are {\it blocked}. 
Below $T_{B}$ at T = 2~K, $M-H$ curves show a wide hysteresis loop for field along $H_\perp$
suggesting a highly oriented nature of the clusters. The Co nanoclusters possess an {\it easy}
and {\it hard} axis of magnetization coupled with the magnetocrystalline anisotropy of 
the Ti$_{1-x}$Co$_{x}$O$_{2}$. In addition at T=2~K, surprisingly a crossover in the magnetization 
for two field directions in sample~B and C is observed. The origin of this crossover is
the anisotropic paramagnetism arising from the 2+ ionic state of Cobalt in a $S=3/2$ system.
Role of dipole- interactions and inter-cluster exchange interactions have also been discussed.

\newpage
\begin{table}[ht!]
\caption {$T_B$ and parameters obtained by fitting $M - H$ data to the 
Langevin function are listed for $H_\perp$ for 
samples A, B and C. $T_B$ for $H_\parallel$ is
also mentioned}  

\label{tab:Tab2} 
\begin {tabular}{|c|c|c|c|c|c|c|}
\hline
Sample& Fluence          &T$_{B}$ (K)& $T_{B}$ (K)    & Particle  &$\sigma$ & Average moment\\
      & (ions/cm$^{2}$)  &($H_\perp$)& ($H_\parallel$)& Size (nm) &         & ${\mu}_{M}$(${\mu}_{B}$)\\\hline
A     &5$\times10^{16}$  &30         &8               &2.50       &0.3      &1261.7\\\hline
B     &8$\times10^{16}$  &65         &40              &3.97       &0.64     &4562.8\\\hline
C     & 1$\times10^{17}$ &150        &70              &4.43       &1.25     &6285.8\\\hline
\end{tabular}
\end{table}

\newpage

\newpage

\newpage

\newpage

\newpage
\begin{figure}
\centering
\includegraphics[width=1.0\textwidth]{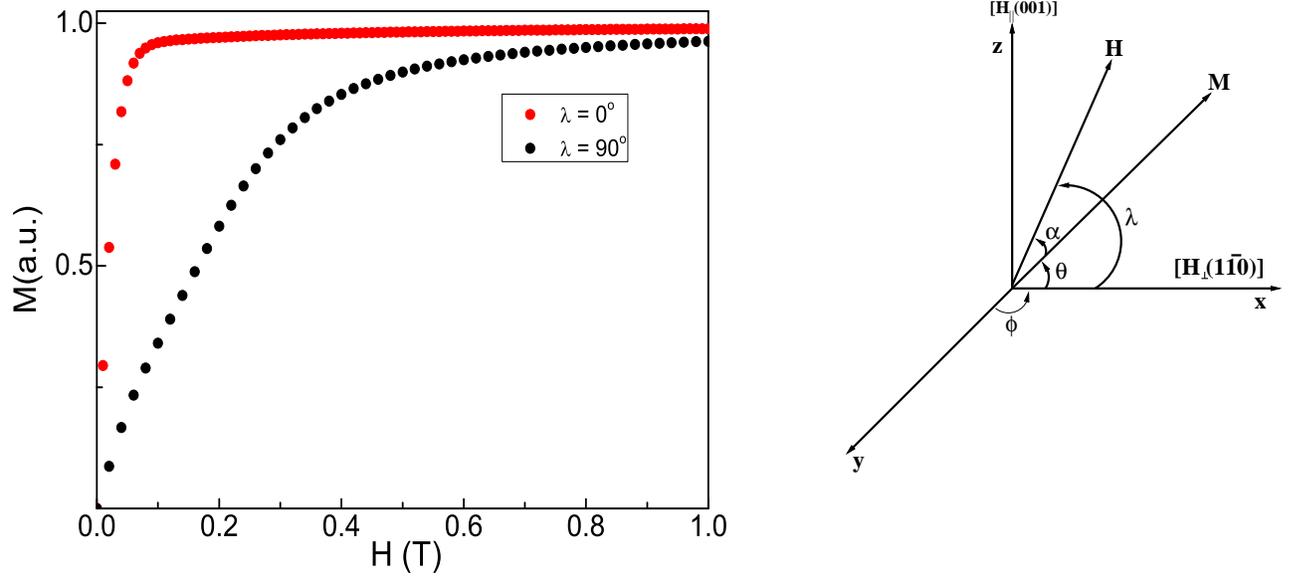}
\caption{$M$-$H$ curves calculated for a single nanocluster
for magnetic field applied along easy axis, ${\lambda}$ = 0$^{o}$ ($H_\perp$) and hard axis, 
${\lambda}$ = 90$^{o}$ ($H_\parallel$) assuming complete saturation of the nano-particle 
with anisotopy constant $K_{mu}$ = 5x10$^{6}$Joules/m$^{3}$.}
\label{fig:fig5}
\end{figure}

\newpage
\begin{figure}
\centering
\includegraphics[width=0.8\textwidth]{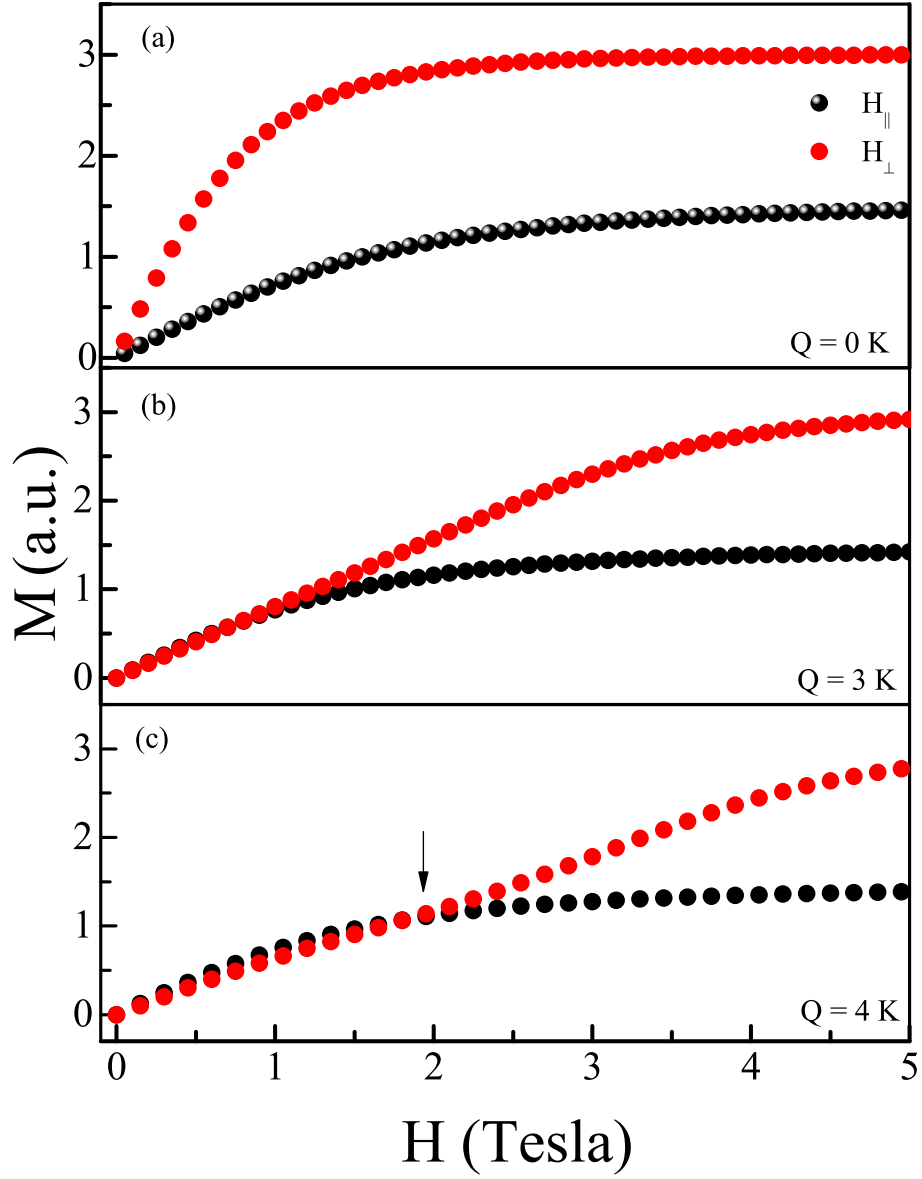}
\caption{Calculated $M$-$H$ curves at $T=2$~K for H $\parallel$ $H_\parallel$
(black symbols) and H $\parallel$ $H_\perp$ (red symbols) using effective spin 
Hamiltonian of eqn.~(4) and 
different values of zero -field splitting $Q$. The anisotropy and the cross-over 
(marked by the arrow) between two field directions observed at $Q = 4$~K are similar 
to experimental results in Fig.~4} 
\label{fig:fig6}
\end{figure}

\newpage
\begin{figure}
\centering
\includegraphics[width=0.85\textwidth]{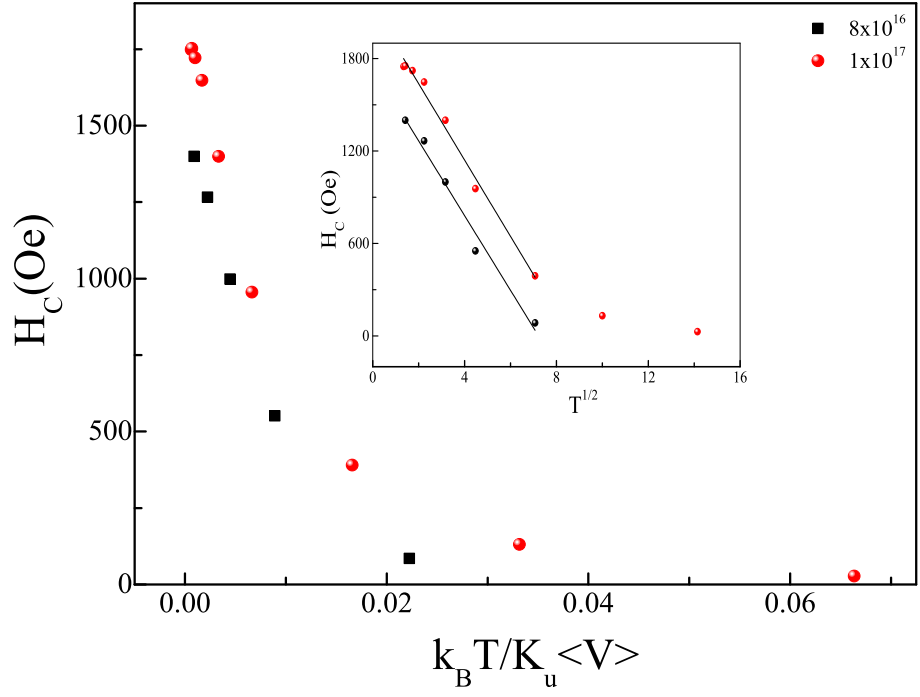}
\caption{$H_{C}$ as a function of the reduced temperature $k_{B}T/K_{u} <V>$, 
for samples~B (square) and C (circle) for the field along the $H_\perp$ direction. Inset 
displays $H_{C}$ as a function of $T^{1/2}$ for samples B and C.}
\label{fig:fig2c}
\end{figure}

\end{document}